\documentclass[pre,aps,twocolumn,floatfix,superscriptaddress]{revtex4}

\usepackage{eso-pic,calc}
\usepackage{graphicx}
\usepackage{amsmath, amsthm, amssymb}
\usepackage{epsfig}
\usepackage{latexsym}
\usepackage{bm}

\def\beqr{\begin{eqnarray}}
\def\eqnr{\end{eqnarray}}
\def\beq{\begin{equation}}
\def\bc{\begin{center}}
\def\ec{\end{center}}
\def\eqn{\end{equation}}

\def\prl#1#2#3{{ Phys. Rev. Lett.} {\bf #1}, #2 (#3)}
\def\epl#1#2#3{{ Euro. Phys. Lett.} {\bf #1}, #2 (#3)}

\def\pre#1#2#3{Phys. Rev. E {\bf #1}, #2 (#3)}

\def\pnas#1#2#3{Proc. Natl. Acad. Sci. (USA) {\bf #1}, #2 (#3)}

\def\physa#1#2#3{Physica A {\bf #1}, #2 (#3)}

\def\sc#1#2#3{Science {\bf #1}, #2 (#3)}

\begin{document}
\title{Critical P{\'o}lya urn}
\author{Avinash Chand Yadav}
\affiliation{Department of Physics and Astronomical Sciences, Central University of Jammu, Samba 181 143, India}

\begin{abstract}
{We propose a variant model of P{\'o}lya urn process,  where the dynamics consist of two competing elements namely, suppression of growth and enhancement of dormant character. Here the level of such features are controlled by an internal parameter in a random manner, so as the system self-organizes and characteristic observables exhibit scale invariance suggesting the existence of criticality.   Varying the internal control parameter, one can explain interesting universality classes for avalanche-type events. We also discuss different versions of the model. It is pointed that such an underlying mechanism has earlier been noted to operate in a neural network. }
\end{abstract}

\maketitle

\section{Introduction}
The P{\'o}lya urn \cite{Johnson_1977, Mahmoud_2008} offers a simple framework for the modelling of  a variety of stochastic processes. Although extensively studied in probability and combinatorics \cite{Feller_1971}, its' applications span across various disciplines and continues to be a topic of active research. Interestingly, urn models have been found to describe diverse phenomena. Examples include spreading of infectious disease, the evolution of species in bioscience \cite{Hoppe_1987}, analysis of algorithms in computer science \cite{Mahmoud_2008}, decision making and reinforced learning \cite{Martina_2002, Pemantle_2007}, and dynamics of novelties \cite{Tria_2014}.

In the standard  P{\'o}lya urn process, the urn consists of two colored balls black and white. A ball may represent a physical entity such as an atom to human or a species to a course of action. At each step, one ball is randomly drawn from the urn and its' color is observed. Then, the ball is returned with a new ball of the same color. This process basically models the {\it rich get richer} phenomenon. 

Antal {\it et. al.} \cite{Antal_2010} have shown that if starting with an imbalanced initial configuration and stopping when a tie occurs for the first time, then the number of balls of one color $L$ satisfies a {\em power law} distribution 
\begin{equation}
P(L) \sim L^{-\tau_L},
\label{prob_l}
\end{equation}
with $\tau_L = 2$. The exit probability that the process ever reaches to a tie is found to be less than one, implying the process is {\em transient}. Note a contrast with the ordinary simple random walk in one dimension: The process is {\em recurrent}, and the first passage time (FPT) exponent has a value 3/2 \cite{Feller_1971, Redner_2001, Bray_2013}.

\begin{figure}[t]
  \centering
  \scalebox{0.25}{\includegraphics{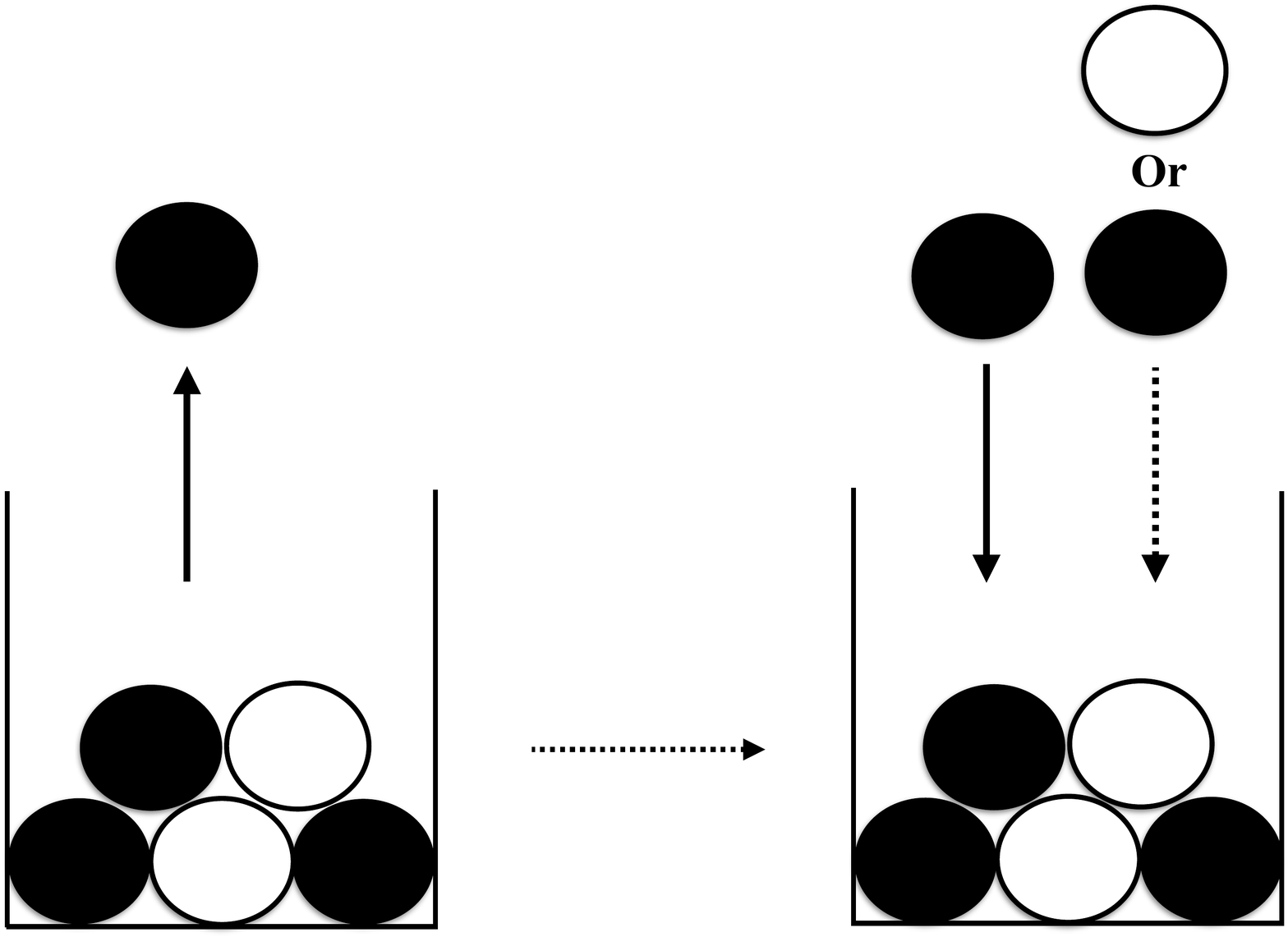}}
  \caption{Sketch of the model  dynamical rule consisting of two steps: (i) A ball is randomly drawn and its color is checked. (ii) The ball is returned with one more ball of the same or of different  color, probabilistically [see Eqs.~(\ref{l_noisy}) and (\ref{I_noisy_pt})]. There are black and white balls in the urn. }
 \label{fig1} 
\end{figure}

It has been observed that the FPT exponent in many systems may take a value different from 3/2 and 2 (see below). Although variants of urn model  offer an explanation for a tunable exponent characterizing a power law distribution for some observables \cite{Tria_2014}, those observables are different from the FPT. Within the framework of urn process, the rich get richer mechanism seems inadequate to explain a broader range of FPT exponent. The purpose of this work is to explore strategies that can explain the aforesaid point.

In this paper, we introduce a variant of P{\'o}lya urn as shown in Fig.~\ref{fig1}, with a specified set of constraints discussed later. The system evolves such that it eventually {\em self-organizes} as a result of   a competition between {\em suppression of growth and enhancement of dormant character}. Also, the system exhibits power law features for characteristic observables, reflecting critical nature.  One can here explain a continuously varying value of the exponent in a range $5/4 \le \tau_L\le 5/2$.  In addition, the process can be mapped to a discrete two dimensional (2-d) random walk that can further be reduced to a one-dimensional (1-d) random walk problem with a suitable transformation. We here provide exact results supported with simulation studies for scaling characteristics of the process.

The exponent in the range $\tau_L\in[3/2, 2]$ can be realized by a 1-d random walk with some additional constraints \cite{Dickman_2001}. For {\em fractional Brownian motion}, the first passage exponent varies as $\tau_L = 2-H$, where the parameter $H$ is the Hurst exponent with $0\le H\le 1$  \cite{Krug_1997}.  The exponent may vary in a range $1< \tau_L \le 5/2$ for diffusion with an absorbing boundary at a marginally moving cliff \cite{Redner_2001}. In the context of avalanches observed in models like  sandpiles \cite{Bak_1996}, the exponent characterizes duration time. For mean field branching process (MFBP), the duration exponent is $\tau_L = 2$ \cite{Zapperi_1995}.  Note that the different exponent values correspond to different {\em universality classes}.

The ubiquitous power-laws or scale invariant features \cite{Newman_2005}, basically emergent behavior of {\em complex systems}, have been of considerable interest, especially in statistical physics.  The {\em central limit theorem} explains the origin of Gaussian statistics, while the emergence  of power law can be understood from a few widely recognized routes such as self-organized criticality \cite{Bak_1987, Yadav_2012}, preferential attachments \cite{Albert_1999}, and sample space reducing (SSR) stochastic processes \cite{Murtra_2015, Yadav_2016_ssr1}.

Sec.~II presents definition of the proposed model, and a mapping with random walk problem. Results for the first passage properties and different versions of the process are also discussed here. Avalanches like events have been studied in Sec.~III, followed by conclusion in Sec.~IV.

\section{Model of critical P{\'o}lya urn}
Consider a P{\'o}lya urn process with two different colored balls, namely, black and white represented instantaneously by a configuration $\{x(t), y(t)\}$. At each elementary step, a ball is selected randomly from the urn and its' color is inspected.  Then the ball is returned with one more ball of the same or different color  with the following stochastic dynamical rules:  
\begin{equation}
\begin{pmatrix} x(t+1) \\ y(t+1) \end{pmatrix}  \to \begin{cases}\begin{pmatrix} x(t)+1 \\ y(t) \end{pmatrix},~{\rm with~rate}~1-p(t),\\ \begin{pmatrix} x(t) \\ y(t)+1 \end{pmatrix},~{\rm with~rate}~p(t).  \end{cases}
\label{l_noisy}
\end{equation}
The time dependent transition rate $p(t)$ is determined as
\begin{equation}
p(t) = \begin{cases}f_{\lambda}[x(t),y(t)],~~~{\rm with~probability}~ \lambda,\\ g_{\lambda}[x(t),y(t)],~~~{\rm with~probability}~1-\lambda, \end{cases}
\label{I_noisy_pt}
\end{equation}
with memory  or strategy functions
\begin{equation}
f_{\lambda}(x,y) = \frac{x}{x+y}~~~{\rm and}~~~g_{\lambda}(x,y) = \frac{1}{2}. 
\end{equation}
Here, $0\le \lambda \le 1$. 
Qualitatively,  the memory dependent transition rate $p(t) = f_{\lambda}$ is chosen with probability $\lambda$, and the memory independent critical transition rate  $p(t) = 1/2$ is selected with the complementary probability $1-\lambda$. Next, a new ball is added of different color or of the same color, with the transition rate $p(t)$ or  $1-p(t)$, respectively. We call the mixed process `critical P{\'o}lya urn-I'.

%The dynamics reflects  a competition between {\em suppression of growth and enhancement of dormant character}. To appreciate the underlying mechanism of the process take $\lambda = 1$, i.e., $p(t)=f_{\lambda}$. When $x>y$ or $y>x$, $p(t)$ or $1-p(t)$ will be dominant, addition of a ball of $y$ or $x$ type will have higher likelihood. 

The strategy functions include  memory dependent or/and independent features. For example, consider a learning process, where the subject forms a decision typically for dichotomous choices, weighing the rewards earned in the past for a course of action \cite{Pemantle_2007}. However, the subject may face occasionally  a conflicting situation due to inevitable reasons. In such a scenario, a decision may be taken in a purely random manner, suppressing the memory. 

\begin{figure}[t]
  \centering
  \scalebox{1.4}{\includegraphics[width=5.5cm, height=4cm]{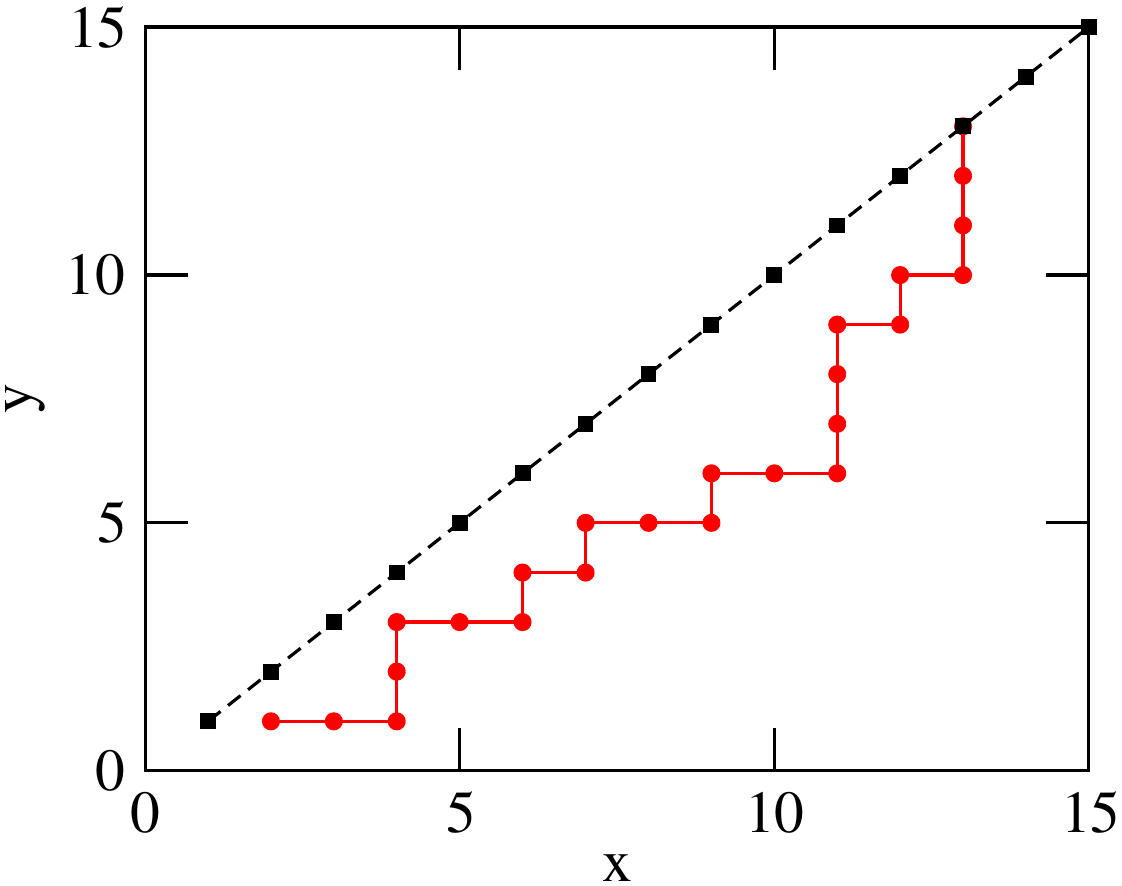}}
   \scalebox{1.4}{\includegraphics[width=5.5cm, height=3cm]{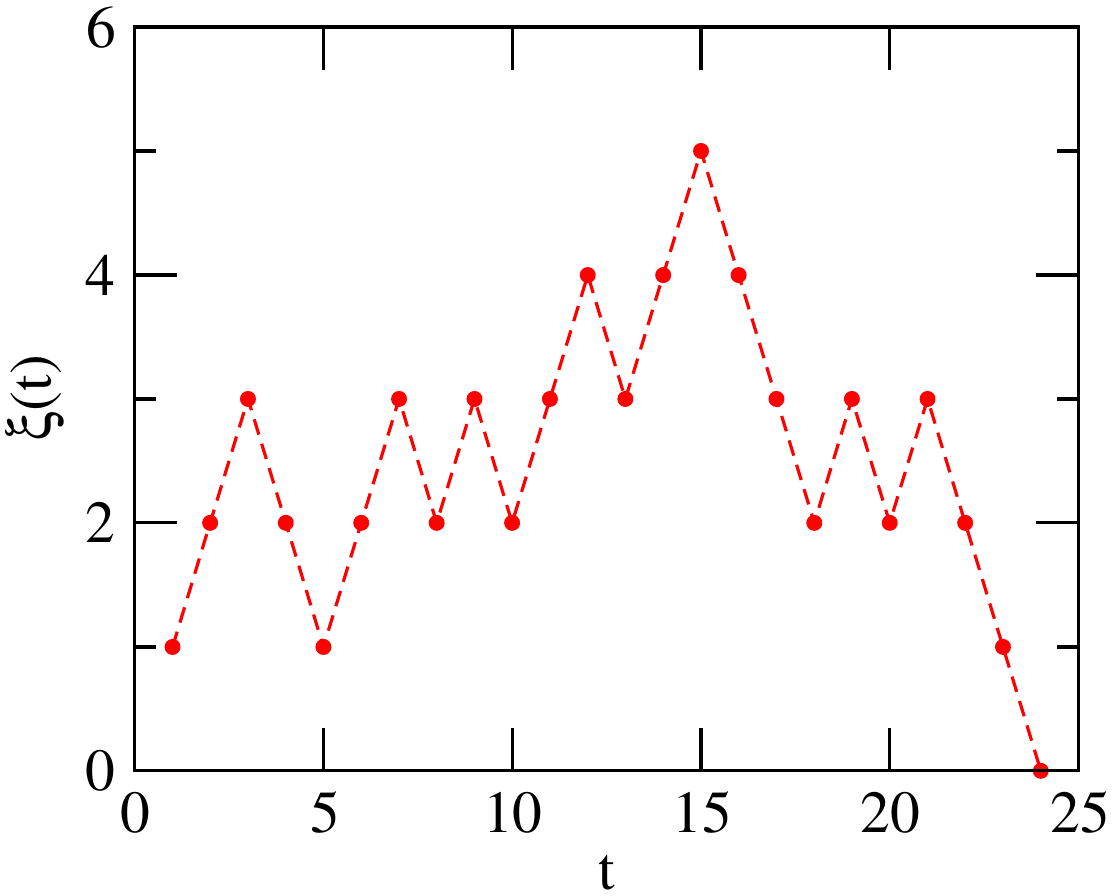}}
  \caption{Top panel: A typical trajectory for the critical P{\'o}lya urn process depicted by staircase function for parameter $\lambda = 1$ as a 2-d random walk. Here, the walk starts from \{2, 1\} and stops at \{13, 13\}, thus $L = 13$ with $T = 23$. Bottom panel: Corresponding 1-d random walk path.}
 \label{fig2} 
\end{figure}

\subsection{A map with random walks}
For this stochastic process, a map with a 2-d random walk with an absorbing boundary condition is clearly evident [see Fig.~\ref{fig2} (top panel)].  The transition probability to move rightward is $p_x$, and a vertical step is taken with the complementary probability $p_y = 1-p_x$. From the definition of the model, the transition probabilities can be easily noted as  
\begin{equation}
p_x = \lambda\frac{y}{x+y} +\frac{1-\lambda}{2}~~~{\rm and}~~~p_y = \lambda\frac{x}{x+y} +\frac{1-\lambda}{2}.
\label{trans_eq}
\end{equation}
The problem can be further reduced to a 1-d discrete random walk with jumps of unit length [see Fig.~\ref{fig2} (bottom panel)], if we introduce a fluctuating quantity $\xi$, namely, the difference in the number of two colored balls as a function of time 
\begin{equation}
\xi(t) = x(t) - y(t).
\label{xi_t}
\end{equation}
Let $\mathcal{P}(\xi;t)$ be  the probability of finding the 1-d random walker at position $\xi$ after time $t$, given the walk starts from $\xi_0$. We find that  $\mathcal{P}(\xi;t)$ satisfies, in the continuum limit, a {\em convection-diffusion} equation of the form [see Appendix~A]
\begin{equation}
\frac{\partial  \mathcal{P}(\xi;t)}{\partial t}- \lambda\frac{\xi}{t} \frac{\partial \mathcal{P}(\xi;t)}{\partial \xi} = \frac{1}{2} \frac{\partial^2 \mathcal{P}(\xi;t)}{\partial \xi^2}.
\label{diff_lam}
\end{equation} 
The case  $\lambda = 0$ or $t\to \infty$  is equivalent to the ordinary 1-d random walk or purely diffusive process.

\subsection{The first passage properties}
The walk starts with a non-zero  imbalanced initial configuration $\{x(0), y(0)\}$ such that either $x>y$ or $y>x$. For example, we take $\{2, 1\}$ throughout in our simulation studies. 
The process is stopped when a tie happens for the first time, i.e., $y(t=T) = x(t=T) = L$.  For the initial condition $\{2,1 \}$, the number of balls of one color $L$ is related to the life span of the process as  $L = (T+3)/2$. Here, an observable of interest may be the time how long the process survives or simply life span $T$ that is a random variable, but an odd number. Alternatively, a more preferable observable is $L$ that is an integer.

\begin{figure}[t]
  \centering
  \scalebox{0.67}{\includegraphics{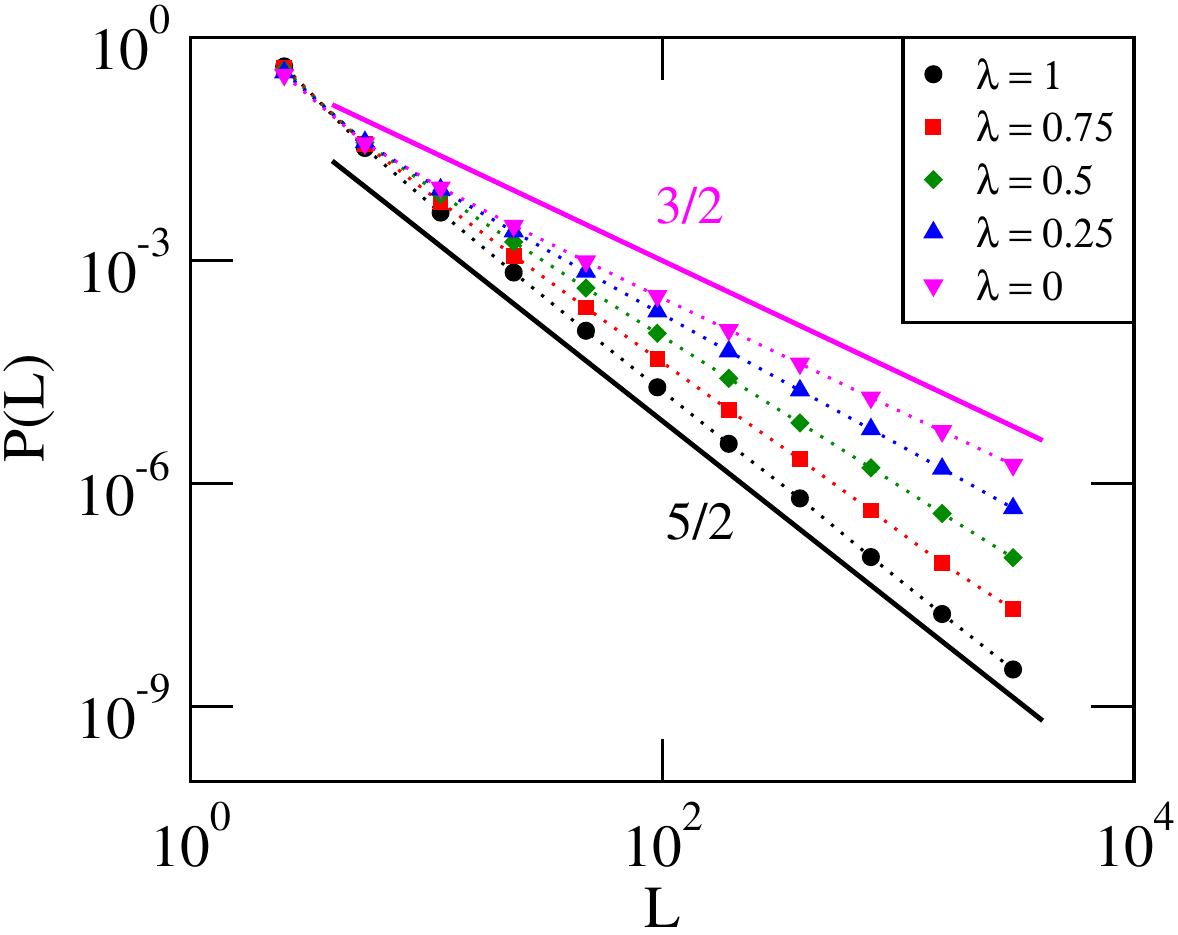}}
  \caption{The distribution $P(L)$ for different values of the parameter $\lambda$. We take $10^8$ independent realizations and the maximum run time for a process is $10^4$. Log-binned data is presented. For comparison, two straight lines are drawn with slopes 3/2 (upper line) and 5/2 (lower line). }
 \label{fig_pl} 
\end{figure}

\begin{figure}[t]
  \centering
  \scalebox{0.67}{\includegraphics{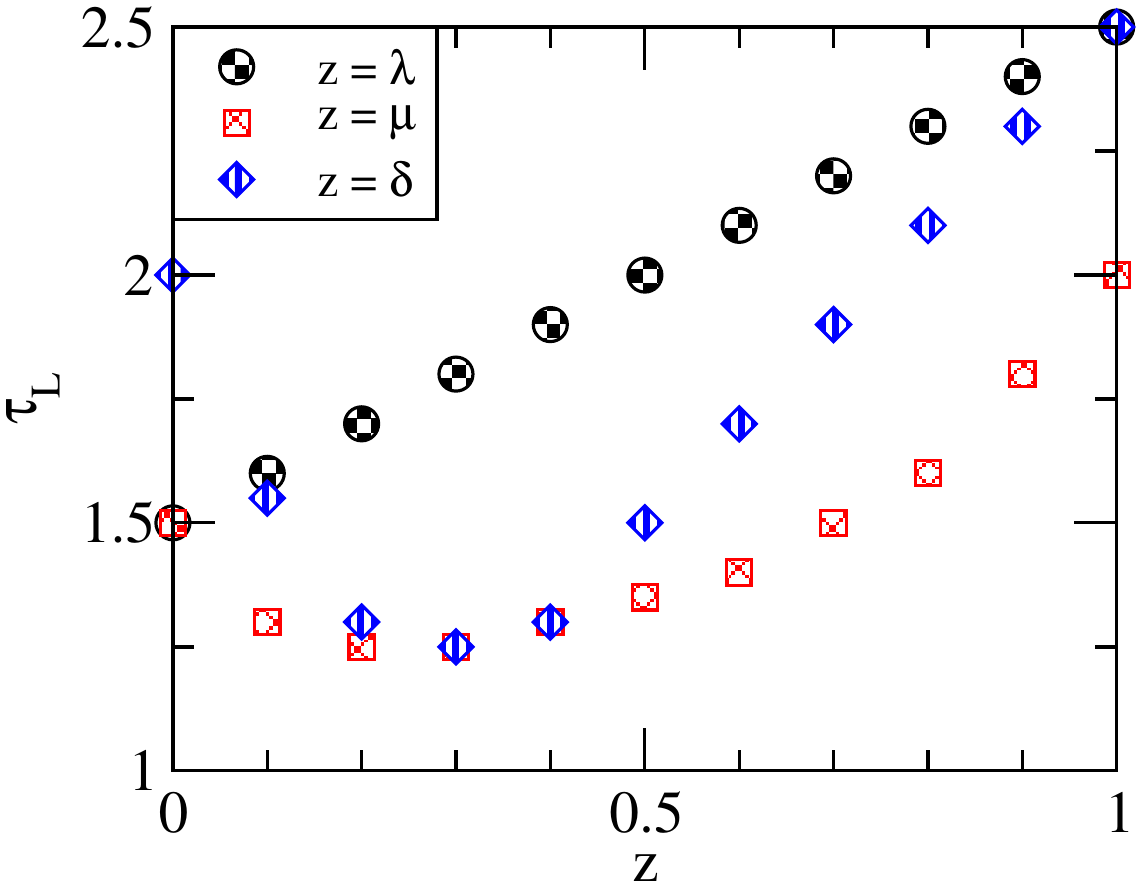}}
  \caption{The  variation of numerically estimated exponent $\tau_L$ with  mixing parameter $z \equiv \lambda, \mu, \delta$ for critical P{\'o}lay urn processes: I, II, III, respectively. Here, $1.25\le \tau_L\le 2.5$.}
  \label{fig_exp} 
\end{figure}

The random walk is non-Markovian, and typically such processes are analytically challenging to track. However, in the present case the problem can be solved, as the velocity [see Eq.~(\ref{vel})] satisfies a simple scaling in state space and time variables. 
In the asymptotic limit, $L\to $ large, the probability distribution $P(L)$ follows a power law form [see Eq.~(\ref{prob_l})]. Our analytical results [see Appendix~A] reveal that the exponent depends upon the mixing parameter $\lambda$  in a simple manner 
\begin{equation}
\tau_L \approx \frac{3}{2} +\lambda,~~~{\rm for}~~0\le \lambda \le 1.
\label{tau_l}
\end{equation}
Eq.~(\ref{tau_l}) suggests that the exponent covers an interesting range $1.5 \le \tau_L \le 2.5$.  Figure~\ref{fig_pl} and \ref{fig_exp} provide a validation of these results obtained by Monte Carlo simulations.

Note that one possible way to explain a tunable exponent for a power law characteristics is by mixing two qualitatively different dynamical elements [as done in Eq.~(\ref{I_noisy_pt})]. For example, in case of scale-free networks the elements are ``popularity'' and ``fitness'' of nodes \cite{Hui_2004}, while for noisy SSR these are ``SSR'' and ``unconstrained hopping'' \cite{Murtra_2015}.

\subsection{Other versions of the process}
 The model discussed above is not unique. One can easily design a modified version with a suitable memory function. 
In the following case, we take the memory functions given as 
\begin{equation}
f_{\mu}(x,y) = \frac{y}{x+y}~~{\rm and}~~g_{\mu}(x,y) = \frac{1}{2}. 
\end{equation} 
Eqs.~(\ref{l_noisy}) \& (\ref{I_noisy_pt}) specify the update rules, with the same initial and boundary conditions. 
Here,  the distribution $P(L)$ still shows a power law, but the exponent $\tau_L$ depends upon $\mu$ in a nonlinear fashion as shown in Fig.~\ref{fig_exp}. We name this version as `critical P{\'o}lya urn-II'. Note that the limiting case $\mu = 1$ corresponds to the  standard urn model.

We can also recognize another version: `critical P{\'o}lya urn-III', by choosing the memory functions as
\begin{equation}
f_{\delta}(x,y) = \frac{x}{x+y}~~{\rm and}~~g_{\delta}(x,y) = 1-f_{\delta}. 
\end{equation}
Then, the configuration $\{x(t), y(t)\}$ is updated following the rules  [Eqs.~(\ref{l_noisy}) \& (\ref{I_noisy_pt})]. In this case, again the exponent $\tau_L$ varies nonlinearly as a function of  $\delta$ (see Fig.~\ref{fig_exp}).

The parameter dependency of the  velocity affects the behavior of the exponent $\tau_L$. Note that the reduced  velocity is $\bar{v}_z = (p_x-p_y)/v$, where  $v = \xi/t$ and $z \equiv \lambda, \mu,$ or $\delta$. For comparison, the reduced velocities are given below:
\begin{equation}
\bar{v}_{\lambda} = -\lambda,~ \bar{v}_{\mu} = \mu,~{\rm and}~\bar{v}_{\delta} = (1-2\delta).
\end{equation} 
Since the parameter $z$ lies between 0 to 1,  the range of reduced velocity is 
\begin{equation}
\bar{v}_{z} \in \begin{cases} [-1, 0],~~{\rm for}~~z = \lambda,\\ [0, 1],~~~~{\rm for}~~z = \mu,\\ [-1, 1],~~{\rm for}~~z = \delta.
\end{cases}
\end{equation}   
For the $\delta$ version of the model, the exponent $\tau_L$  can be determined by directly comparing the reduced velocities of the $\lambda$ and $\mu$ versions. %When $1/2 \le \delta \le 1$, the reduced velocity is negative i.e., $\bar{v}_{\delta}  \in [-1, 0]$. Consequently, the exponent is linear  similar to that of $\lambda$ version model. When $0 \le \delta < 1/2$, the reduced velocity is positive i.e., $\bar{v}_{\delta}  \in [0, 1]$ and  the exponent behaves nonlinearly similar to that of $\mu$ version of the model.  
Mathematically, we get
\begin{equation}
\tau_L(\delta) = \begin{cases}\frac{3}{2} + (2\delta-1),~~~~~{\rm for~~}1/2 \le \delta \le 1,\\ \tau_L(\mu = 1-2\delta),~~{\rm for~~}0 \le \delta < 1/2.   \end{cases}
\end{equation}

\section{Avalanches like events}
\begin{figure}[t]
  \centering
  \scalebox{0.65}{\includegraphics{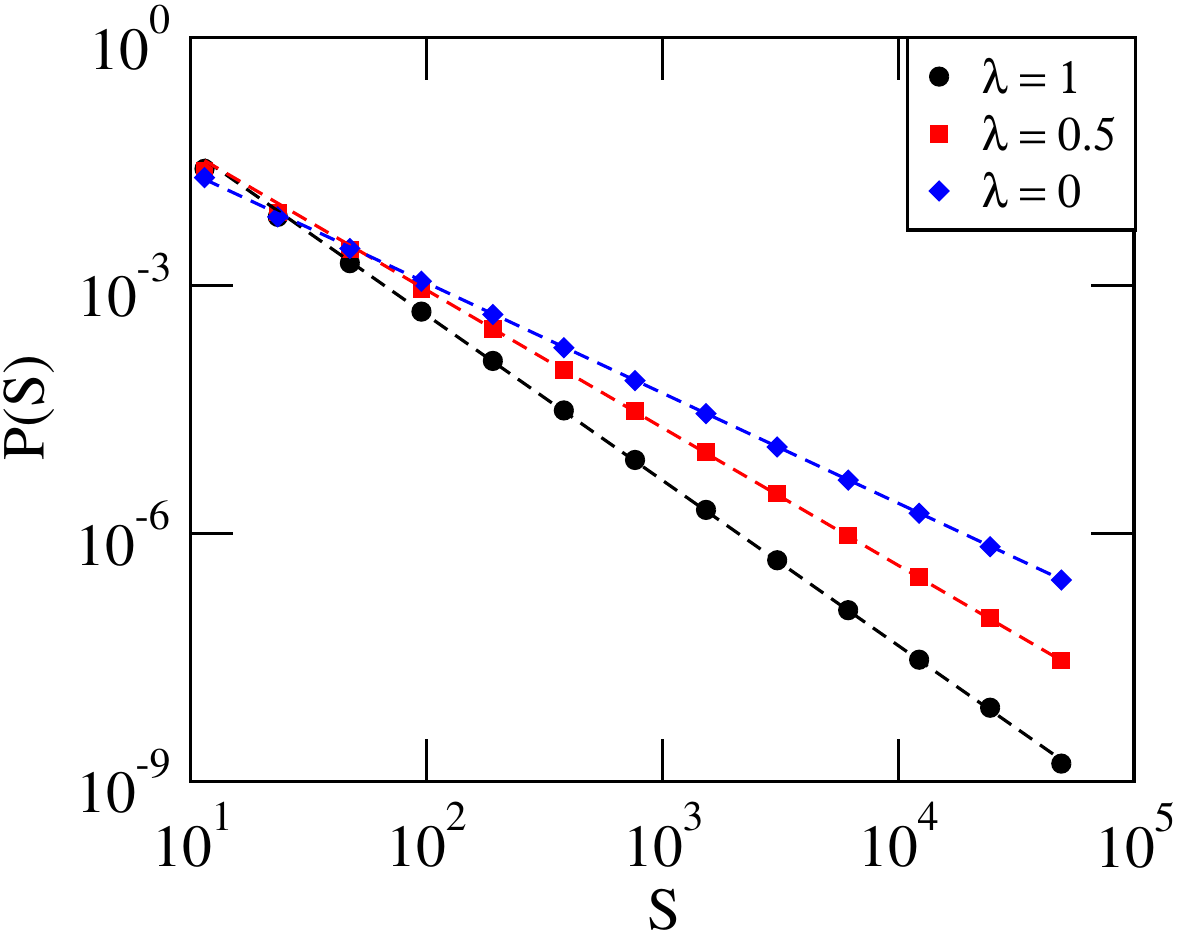}}
  \scalebox{0.65}{\includegraphics{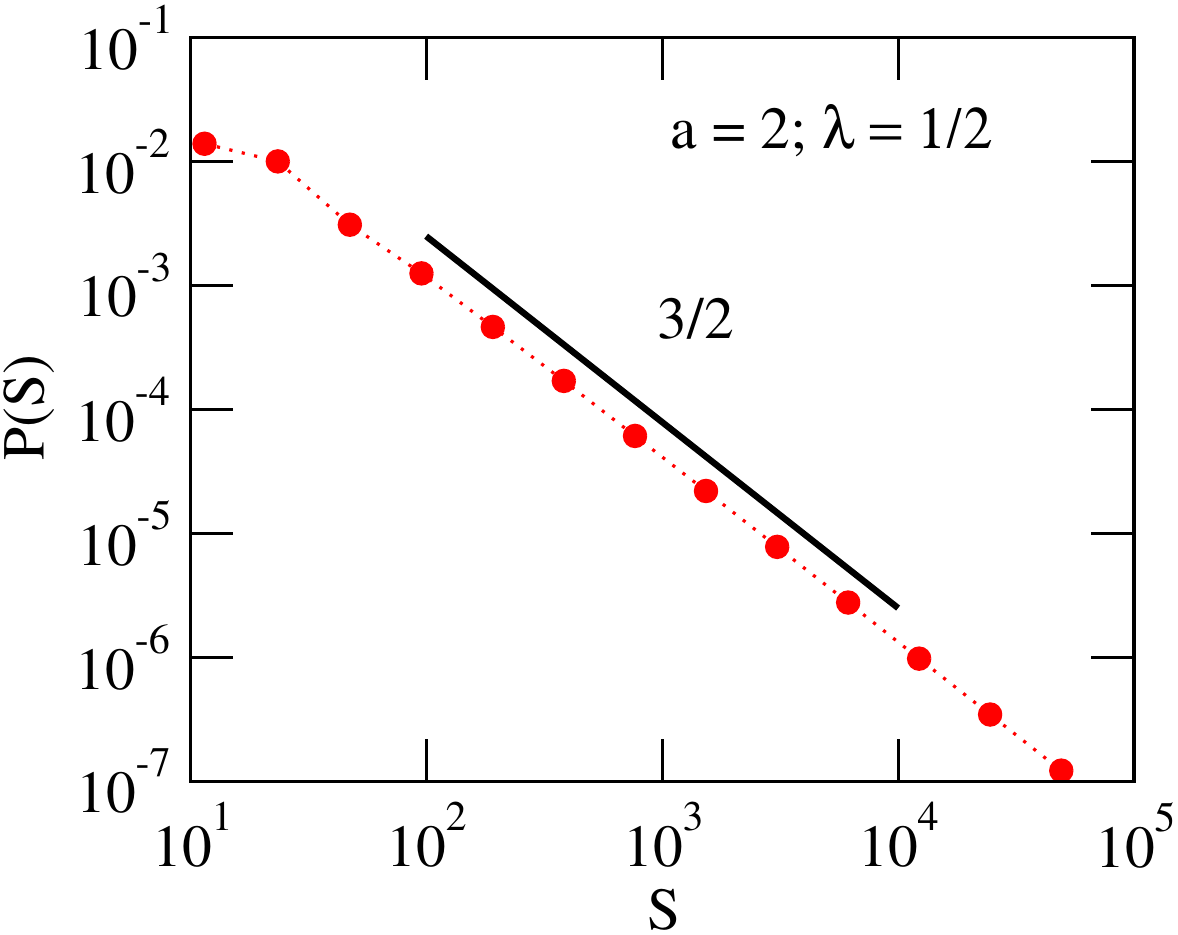}}
  \caption{Top panel: The cluster size distribution $P(S)$ for $a = 1$ and different values of  parameter $\lambda $, along with the best fit curves shown by dashed lines. The slopes are $\tau_S = $ 2, 5/3, 4/3 for $\lambda = 1$, 0.5, and 0, respectively. Bottom panel: The plot of $P(S)$ for  parameters $a = 2$ and $\lambda  = 1/2$.  A straight line with slope 3/2 is drawn for comparison.}
 \label{fig_ps_2} 
\end{figure}

Recall the activity signal [see Eq.~(\ref{xi_t})]. This transformation results in a 1-d random walk with only positive positions with initial setting $x>y$. Note that such a difference process is subtly more informative similar to that  of incremental values in the case of records \cite{Schehr_2016}. One can further characterize $\xi(t)$ with {\em avalanche} like observables  such as size $S = \sum_{t=0}^{T}\xi(t)$ and maximum activity $M = {\rm max}\{\xi(t)\}$.   
For {\em critical avalanches}, observable $X\in \{L, S, M\} $ is a random variable characterized by probability distribution of the power law type $P(X) \sim X^{-\tau_X}$. Moreover, any two observables are related as $\langle X\rangle \sim Y^{\gamma_{XY}}$. The different exponents are connected as 
$\tau_X = 1+ [(\tau_Y-1)/\gamma_{XY}],$ 
and $\gamma_{XY} = 1/\gamma_{YX}$ \cite{Manchanda_2013}.  
Numerically, we observed that the exponent $\gamma_{SL}$ is equal to 3/2 and independent of $\lambda$. 
Consequently, the size exponent is
\begin{equation}
\tau_S = 1+\frac{1+2\lambda}{3}.
\label{ps_exp}
\end{equation}
Figure~\ref{fig_ps_2} (top panel) shows the simulation results for the probability distribution $P(S)$ with different values of the parameter $\lambda$. An excellent agreement with Eq.~(\ref{ps_exp}) is seen.

%\subsection{Universality classes}
Note that $\tau_L = 2$ and $\tau_S = 5/3$ for  $\lambda = 1/2$ [also $\mu = 1$ and $\delta = 3/4$]. Their  universality class is different from that of the MFBP  for which $\tau_S = 3/2$, although $\tau_L = 2$.
As the avalanche type events are nonlinear response of the system, the above definition for the cluster size seems inadequate. We assume that $S = \sum_{t=0}^{T}\xi^{a}(t) $, where $a$ is an index coding the degree of nonlinearity. For $a = 2$, using the relations $\langle \xi^2\rangle \sim t$ (see below), we find 
\begin{equation}
\langle S \rangle = \int_{t=0}^{T}\langle \xi^2\rangle dt \sim \int_{l=0}^{L} ldl \sim L^2.\nonumber
 \end{equation}
 Here, the size exponent is modified as 
\begin{equation}
 \tau_S = 1+\frac{1+2\lambda}{4}~~~~~{\rm~for~}a = 2.
 \end{equation}
Thus, the case  $\lambda = 1/2$ yields $\tau_S = 3/2$ [see Fig.~\ref{fig_ps_2} (bottom panel)] and  corresponds to MFBP universality class.  Note that $\tau_L = 3/2$ and $\tau_S = 4/3$ with $a = 1$ for $\lambda = 0$ [also $\mu = 0.75$ and $\delta = 1/2$], and theses cases belong to a universality class related to directed sandpile model introduced by Dhar \& Ramaswamy \cite{DR_1989}.

\begin{figure}[t]
  \centering
   \scalebox{0.58}{\includegraphics{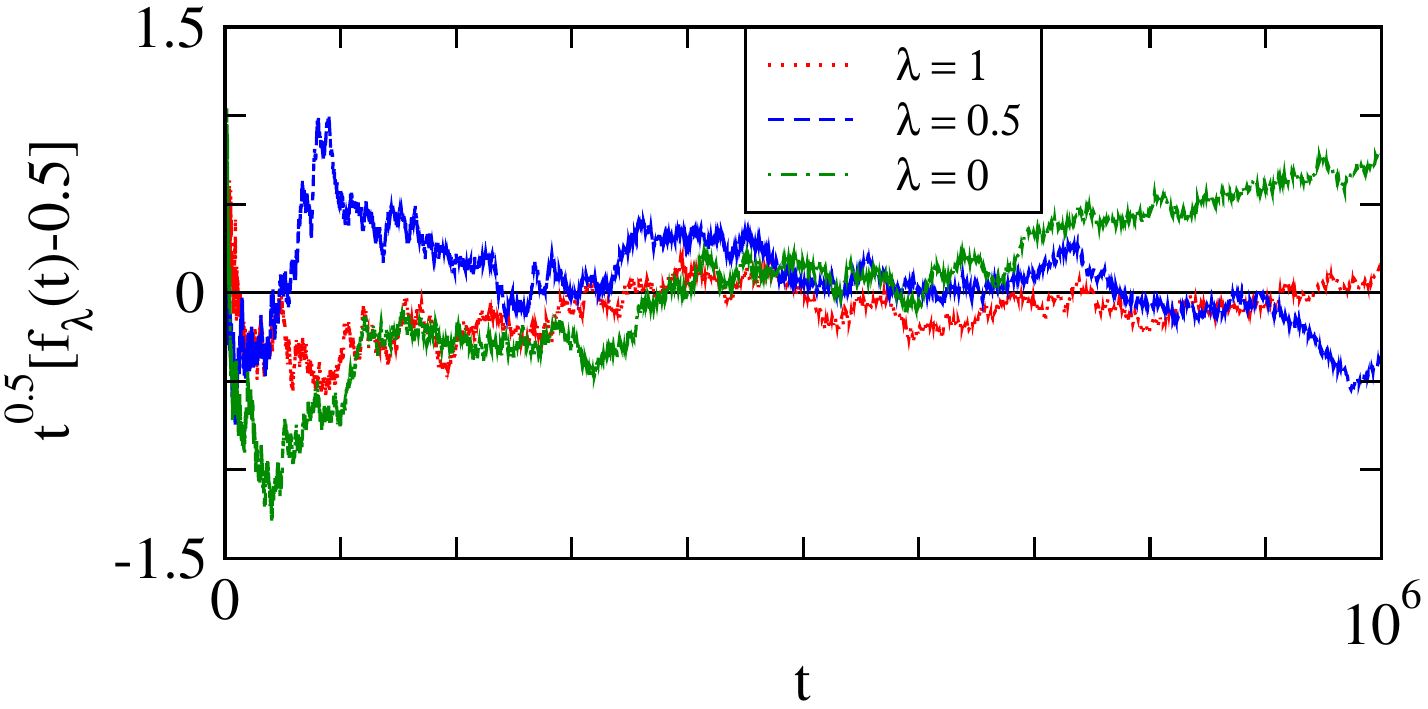}}
    \caption{Signature of self-organization: A typical plot of the scaled indicator function $\sqrt{t}[f_{\lambda}(t)-0.5]$. }
 \label{fig_so} 
\end{figure}

%{\bf Self-organization.--}
Here, a  negative feedback mechanism operates to keep $\xi(t)$ tending to 0, randomly. 
When an indicator function is such that $f_{\lambda}(t)>1/2$, we get $\xi(t)>0$. To bring the $\xi(t)$ towards 0 randomly, a ball of $y$ type would be preferred to be added so that the indicator function can be shifted towards 1/2.  The self-organization occurs bringing towards critical state for this non-equilibrium system characterized by events exhibiting scale-free statistics or the existence of long range correlations \cite{Sornette_2006}.

To explore how the system self-organizes, we study evolution of the fluctuating quantity $f_{\lambda}(t)$. The process starts with an imbalanced configuration and runs for a long time. If $x = y$, then $f_{\lambda} = 1/2$. We find that $\Delta f_{\lambda} = f_{\lambda}(t)-0.5$ hovers about 0, but the fluctuations decrease on increasing the observation time [see Fig.~\ref{fig_so}],
\begin{equation}
\langle \Delta  f_{\lambda}(t)\rangle = 0~~{\rm and}~~ \sqrt{\langle [\Delta  f_{\lambda}(t)]^2\rangle} \sim 1/\sqrt{t}.
\end{equation} 
Also, $\xi(t)$ hovers about 0, but as the random walk drifts, the fluctuations increase with increasing the observation time \begin{equation}
\langle \xi(t)\rangle = 0~~{\rm and}~~\sqrt{\langle \xi^2\rangle} \sim \sqrt{t}. 
\end{equation}

Clearly, except for the case $\lambda = 0$ where $p(t) = 1/2$, in all other cases there is no tuning of an external control parameter. The systems exhibit criticality that emerges as a consequence of self-organization that results from a competition between  suppressing growth and enhancing dormant character in the system.

Such an underlying mechanism has earlier been  noted by Bornholdt and Rohlf \cite{Bornholdt_2000}  in a random network with binary elements. A local rewiring rule is used: A node that is active, over an attracting period, loses a link and a quiet node gets a new link. Eventually this leads to criticality, emerging as a result of  self-organization, with an application to  neural network explaining how the density of connections is regulated. Since activity-dependent attachment of synapses  to a neuron is responsible for that and remains an experimentally observed  feature \cite{Rauschecke_1979}.

\section{Conclusion} We have introduced a class of P{\'o}lya urn process that exhibits scaling features. The underlying mechanism is the competition between suppression of growth and enhancement of dormant character. The process can be eventually mapped to a 1-d discrete random walk with an absorbing boundary condition. The first passage properties show a remarkably rich behavior, exhibiting from Gaussian to power law type statistics with the duration exponent continuously varying between 5/4 to 5/2. The value of the exponent is determined by an internal parameter controlling  the rate how the memory dependent transition probability is preferred.  The exponent range is reasonably wide and physically quite interesting.  We have also discussed the characteristics of critical avalanches.

%{\color{blue}
Interestingly,  a variant of P{\'o}lya urn also offers a deeper understanding of Elephant Random Walk (ERW) \cite{Trimper_2004} that is a 1-d discrete time random walk with a complete memory of its past \cite{Baur_2016}. The ERW basically explains anomalous diffusion, and is a class of explicitly solvable non-Markovian  random walk. At each time step, the ERW moves  by randomly selecting a step taken at a previous time,   with a probability $p$. Otherwise,  it will take an opposite step with the complementary probability. In the urn scheme, a ball is randomly selected from the urn and its color is checked. Then the ball is returned with a new ball of the same color with probability $p$ and a ball of different color with the complementary probability $1-p$. The difference between the number of two colored balls represents the position of the ERW. Variants and extension to higher dimensions of ERW can also be described using the urn framework \cite{Baur_2016}. We note an important difference between our urn models and the corresponding urn for ERWs that in the former case the evolution depends  upon a memory function that  includes a state space dependent and/or independent features.

The standard P{\'o}lya urn process can be generalized in a number of ways: For examples, considering multiple colored balls, adding more than one new ball of same or different color, and/or multiple interacting urn, etc. This flexibility has been used to design different variants. It would be interesting to explore the strategies and the extent of scaling behavior of our model in other variants of the urn process.

\section*{ACKNOWLEDGMENT}
ACY is greatly benefited from R. Ramaswamy with a stimulating discussion on this topic, and would also like to acknowledge support from Department of Science \& Technology, India through a grant ECR/2017/001702.

\appendix
\section{Master equation approach}
To approach the problem analytically, we consider the 2-d random walk representation of the model. We here discuss only the $\lambda$ version of the model in detail, as in other cases the same approach can easily be extended.  Let $\mathcal{P}(x,y;t)$ denotes the probability of finding the 2-d random walker at position $\{x,y\}$ after time $t$, given the walk starts from $\{x_0, y_0\}$   and finally  stops when $x(t=T) = y(t=T) = L$. The master equation reads
\begin{equation}
\mathcal{P}(x,y;t+1) =  p_x\mathcal{P}(x-1,y;t) +  p_y\mathcal{P}(x,y-1;t),
\label{master_eq}
\end{equation}
where $p_x$ and $p_y$ are the  transition probabilities to move by unit step in rightward (along $x$ axis) and upward (along $y$ axis) directions, respectively. Explicitly, the  transition probabilities are expressed as 
\begin{equation}
p_x = \lambda\frac{y}{x+y} +\frac{1-\lambda}{2}~~~{\rm and}~~~p_y = \lambda\frac{x}{x+y} +\frac{1-\lambda}{2},
\label{trans_eq1}
\end{equation}
with $p_x+p_y = 1$.
The transition probability $p_x$ or $p_y$ is  a space and time dependent function, although the time dependency comes in an implicit manner. Thus, the process is non-Markovian or history dependent. Consequently, solving such a master equation seems quite challenging.

Nevertheless, if one considers the difference process 
\begin{equation}
\xi(t) = x(t)-y(t),
\end{equation}
 the problem reduces to a 1-d random walk. Note that 
\begin{equation}
\xi(t+1) = \xi(t) +\eta(t),
\end{equation}
where $\eta$ is a random variable that takes two values $ \pm 1$ with probabilities $p_+(t) = p_x$ and $p_-(t) = 1-p_+(t)$, respectively. 
Thus, the Eq.~(\ref{master_eq}) reduces to
\begin{equation}
\mathcal{P}(\xi;t+1) =p_+\mathcal{P}(\xi-1;t) +  p_-\mathcal{P}(\xi+1;t),
\label{master_xi}
\end{equation}
with an absorbing boundary condition at $\xi = 0$. The time dependency of  $p_+(t)$ and $p_-(t)$ is explicitly given below.

As the time elapsed  is equal to the  total number of balls in the urn, we have $t \approx x+y$. One can easily write $x$ and $y$ in terms of $\xi$ and $t$ as 
\begin{equation}
x = \frac{t+\xi}{2}~~~{\rm and}~~~ y = \frac{t-\xi}{2}.
\end{equation} 
Then, the Eq.~(\ref{trans_eq1}) eventually reads
\begin{equation}
p_{\pm}(t) = \frac{1}{2}\left[1\mp \lambda\frac{\xi}{t}\right].
\label{trans_plus}
\end{equation}

Plugging Eq.~(\ref{trans_plus}) into Eq.~(\ref{master_xi}), we get 
\begin{eqnarray}
\mathcal{P}(\xi;t+1) =\frac{1}{2}[\mathcal{P}(\xi-1;t) + \mathcal{P}(\xi+1;t)]  \nonumber\\+\lambda\frac{\xi}{2t}[\mathcal{P}(\xi+1;t)  - \mathcal{P}(\xi-1;t)].
\label{master_xi_2}
\end{eqnarray} 
In the continuum limit, this reduces to a partial differential equation 
\begin{equation}
\frac{\partial  \mathcal{P}(\xi;t)}{\partial t}- \lambda\frac{\xi}{t} \frac{\partial \mathcal{P}(\xi;t)}{\partial \xi} = \frac{1}{2} \frac{\partial^2 \mathcal{P}(\xi;t)}{\partial \xi^2}.
\label{diff_lam_1}
\end{equation}
This is basically a convection-diffusion equation with the boundary condition $\mathcal{P}(\xi=0;t) = 0$, describing biased random walk with the diffusion coefficient 1/2 and  the velocity given as
\begin{equation}
v_{\lambda} = -\lambda\frac{\xi}{t} = p_+-p_-.
\label{vel}
\end{equation}
Such a convection-diffusion equation appears in a class of diffusion with an absorbing boundary condition at a marginally moving cliff \cite{Redner_2001}.

A comparison with ordinary random walk for which the diffusion equation has the form
\begin{equation}
\frac{\partial  \mathcal{P}(\xi;t)}{\partial t} = \frac{1}{2} \frac{\partial^2 \mathcal{P}(\xi;t)}{\partial \xi^2},
\label{meq_rw}
\end{equation}
 suggests that the  Eq.~(\ref{diff_lam_1}) has an additional term with a prefactor $\lambda(\xi/t)$. This term is sufficiently smaller than 1, because $0\le \lambda \le 1$ and $\xi/t \ll 1$ [as $\langle \xi^2 \rangle \sim t$]. The long time behavior of Eq.~(\ref{diff_lam_1}) is also same as that of $\lambda = 0$.

The solution of Eq.~(\ref{meq_rw}) is the Gaussian function. One can analytically study the survival and eventually the first return time probability distributions. The survival probability varies as 
\begin{equation}
\mathcal{S}(t) \sim t^{-\frac{1}{2}}.
\end{equation}
With the survival probability, it is easy to calculate the first passage probability as these two are related
\begin{equation}
P(L) = -\frac{d\mathcal{S}(t)}{dt}|_{t=T\sim L}.
\end{equation}
In the large limit of $L$, the first passage probability behaves as  
\begin{equation}
P(L) \sim L^{-3/2}.
\end{equation}

\subsection{First passage probability for $\lambda\neq 0$}
In order to solve  Eq.~(\ref{diff_lam_1}) introduce a dimensionless variable $u = [\xi/\sqrt{t}]-1$, and make use of  the scaling ansatz  for the occupation probability $\mathcal{P}(\xi;t)$  as
\begin{equation}
\mathcal{P}(\xi;t) = t^{-\tau-\frac{1}{2}}\mathcal{C}(u).
\end{equation}
The exponent of time dependent prefactor is chosen in such a manner so that the survival time probability should vary as $t^{-\tau}$. Note that $\tau =  \tau_L-1$.
Then, we get an ordinary  differential equation 
\begin{equation}
\frac{1}{2}\frac{d^2\mathcal{C}}{d u^2} + \lambda (u+1) \frac{d\mathcal{C}}{d u} + \left(\tau+\frac{1}{2}\right)\mathcal{C} = 0.
\end{equation}
To further simplify the above equation, it is convenient to introduce ${\bf u} = \sqrt{2\lambda}(u+1)$ and $\mathcal{D}({\bf u}) = \mathcal{C}(u)\exp({\bf u}^2/4)$. This transformation leads to 
\begin{equation}
\frac{d^2\mathcal{D}}{d {\bf u}^2} +  \left[{\rho} +\frac{1}{2} -\frac{{\bf u}^2}{4}\right]\mathcal{D} = 0,~~~{\rm~with}~~{\bf \rho} = \frac{\tau}{\lambda}+\frac{1}{2\lambda}-1.\nonumber
\end{equation}
This is the ordinary differential equation satisfying parabolic cylinder function $\mathcal{D}_{\rho}({\bf u})$. Applying the boundary condition on this function, the first passage exponent $\tau_L$ as a function of $\lambda$ can be estimated. 
Then, we have
\begin{equation}
\mathcal{D}_{\rho}(\sqrt{2\lambda}) = 0.
\end{equation}

The parabolic cylinder function with real argument reads 
\begin{equation}
\mathcal{D}_{\rho}(z) = 2^{\frac{\rho}{2}}\exp\left(-\frac{z^2}{4}\right)\mathcal{U}\left(-\frac{\rho}{2},\frac{1}{2}, \frac{z^2}{2}\right),
\label{pcf}
\end{equation} 
where $\mathcal{U}(\cdot)$ is the confluent hypergeometric function of the first kind. 
In our problem 
\begin{equation}
\rho = \frac{1}{\lambda}\left[\tau_{L}-1+\frac{1}{2}\right] -1,
\end{equation} 
 and $z = \sqrt{2\lambda}$. Here the parameter $\lambda$ is  a positive real number, implying $z$ is also real. The boundary condition   $\mathcal{D}_{\rho}(z) = 0$ suggests that $\mathcal{U}(-\rho/2,1/2,z^2/2) = 0$, since the prefactors in Eq.~(\ref{pcf}) cannot be zero. The confluent hypergeometric function of the first kind is given by 
\begin{equation}
\mathcal{U}(m,n,\lambda) = 1 + \frac{m}{n}\lambda+ \frac{m}{n}\frac{(m+n)}{(n+1)}\frac{\lambda^2}{2} + \dots.
\end{equation} 
Since the parameter $\lambda$ is less than or equal to 1, the higher order terms of $\lambda$ would be small. Using the boundary condition and dropping the higher order terms, we have
\begin{equation}
0 \approx 1 + \frac{-\rho/2}{1/2}\lambda.
\end{equation} 
Simplifying this, we get  the main result
\begin{equation}
\tau_{L} \approx \frac{3}{2} + \lambda,
\end{equation} 
the leading order behavior of the exponent as a function of $\lambda$. 

\begin{figure}[t]
  \centering
  \scalebox{0.65}{\includegraphics{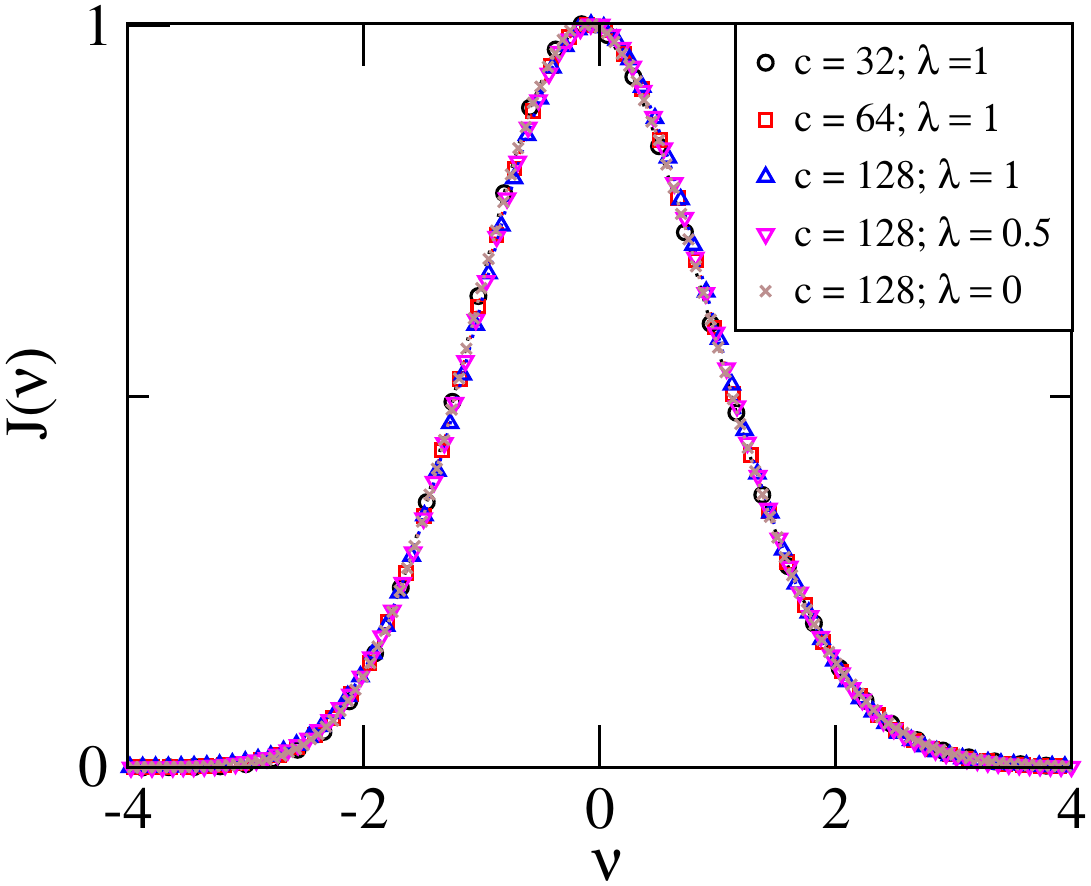}}
  \caption{Data collapse curves for the distribution $P(L)$ with different values of $c$ and $\lambda$.   Here $\nu = (L-\langle L\rangle)/\sigma_L$ and $J(\nu) = \sqrt{2\pi \sigma^{2}_{L}}P(L)$.}
 \label{fig_gauss} 
\end{figure}

\subsection{Key features for other versions}

For $\mu$ and $\delta$ versions, the transition probabilities are explicitly given below: 
\begin{eqnarray}
p_x = \mu\frac{x}{x+y} +\frac{1-\mu}{2}, \nonumber\\ p_y = \mu\frac{y}{x+y} +\frac{1-\mu}{2},
\end{eqnarray}
and
\begin{eqnarray}
p_x = \delta\frac{y}{x+y} +(1-\delta)\frac{x}{x+y},\nonumber\\ p_y = \delta\frac{x}{x+y} +(1-\delta)\frac{y}{x+y}.
\end{eqnarray}
Using Eq.~(\ref{vel}), the  velocity  can be easily computed for each cases as
\begin{equation}
v_{\lambda} = -\lambda\frac{\xi}{t},~~ v_{\mu} = \mu\frac{\xi}{t},~~{\rm and}~~v_{\delta} = (1-2\delta)\frac{\xi}{t}.
\end{equation}

\section{Condition for Gaussian statistics}
Interestingly, the distribution $P(L)$ exhibits Gaussian form [see Fig.~\ref{fig_gauss}], if $y(t=T) = c$ where $c$ is a constant. We numerically find that the mean and variance of $L$ grow with increasing $c$ at constant $\lambda$, while the mean is constant but the variance decreases when $\lambda$ is increased at  fixed $c$.

\end{document}